\documentclass[12pt]{article}
\usepackage{braket}
\usepackage{amssymb}
\usepackage{amsmath}
\usepackage{youngtab}
\usepackage{cite}
\usepackage[]{hyperref}
\usepackage{multirow}
\input xy
\xyoption{all}

\newcounter{daggerfootnote}
\newcommand*{\daggerfootnote}[1]{%
	\setcounter{daggerfootnote}{\value{footnote}}%
	\renewcommand*{\thefootnote}{\fnsymbol{footnote}}%
	\footnote[2]{#1}%
	\setcounter{footnote}{\value{daggerfootnote}}%
	\renewcommand*{\thefootnote}{\arabic{footnote}}%
}

\numberwithin{equation}{section}

\begin{document}

\begin{center}

{\large\bf A Mathematica code for calculating massless spectrum of \\ (0,2) Landau-Ginzburg orbifold}

\vspace{0.2in}

Hadi Parsian\daggerfootnote{Corresponding author}, He Feng 

Department of Physics\\
Virginia Tech\\
850 West Campus Dr.\\
Blacksburg, VA  24061\\
{\tt varzi61@vt.edu}, {\tt fenghe@vt.edu}

$\,$

\end{center}

In this short paper, we try to explain how to use our program which has been written in Wolfram Mathematica to get the massless spectrum of any Landau-Ginzburg orbifold. The technique has been developed by Witten-Kachru theoretically, but calculating it for an explicit Landau-Ginzburg model is exhausting and in general, beyond human ability to calculate using pen and paper.

\begin{flushleft}
January 2020
\end{flushleft}

\newpage

\tableofcontents

\newpage

\section{Introduction}

Landau-Ginzburg orbifolds can be derived from low energy limit of Gauged Linear Sigma models, as their non-geometric phase. They can be used for compactification of Heterotic string theory. There is an advantage in using Landau-Ginzburg models over Calabi-Yau sigma models: one can compute massless spectrum of Calabi-Yau sigma models by computing suitable cohomology groups, but then, the instanton corrections might give mass to some of these states. For Landau-Ginzburg models, as has been shown in \cite{Kachru:1993pg}, the computation is exact. 

For phenomenological and also computational reasons, we have concentrated on Landau-Ginzburg orbifold with central charges $(c,\bar{c})=(6+r,9)$ in our program. In section two, we explain about the model and in section three we will explain how to use our program.

\section{Background and Methods} 
In this section we review the theoretical background to find massless spectrum of $(0,2)$ Landau-Ginzburg orbifolds \cite{Kachru:1993pg,Distler:1993mk}. 
A $(0,2)$ Landau-Ginzburg is formulated in $(0,2)$ supersymmetry language. We have two classes of superfields: chiral superfields and fermi superfields. A chiral superfield, $\Phi$, can be expanded to 
\begin{equation}
	\Phi=\phi+\theta\psi+\theta\bar{\theta}\partial_{\bar{z}}\phi
\end{equation} 
where $\phi$ is a bosonic and $\psi$ is fermionic right-moving field. A fermi superfield, $\Lambda$, can be expanded to 
\begin{equation}
	\Lambda=\lambda+\theta l +\theta\bar{\theta}\partial_{\bar{z}}\lambda
\end{equation}
where $\lambda$ is a left-moving fermionic field and $l$ is an auxiliary field. 
Using this formulation, we can write our (0,2) Landau-Ginzburg theory in the following form:
$$S_{\mathcal{W}}=\int d^2zd\theta \mathcal{W} \quad +\quad \text{c.c.} $$
where $\mathcal{W}$ is called the superpotential of the (0,2) Landau-Ginzburg theory and it has the following general form:
$$\mathcal{W}=\sum_{a} \Lambda^{a} F_{a}(\Phi_i) $$ 
where $\Lambda^{a}$ are Fermi superfields, $\Phi_i$ are chiral superfields, and $F_a$ are holomorphic functions of the chiral superfields. 

In an standard way and for practical reason, to get a superconformal field theory with central charges $(c,\bar{c})=(6+\tilde{r},9)$, we attach $N=16-2\tilde{r}$ free left-moving fermions 
$$L^{\prime}=\int\sum_{I=1}^{N} \lambda^I \bar{\partial}\lambda^I,$$ 
which generate an $SO(N)$ current algebra. This group combines with the left-moving $U(1)$ charges of the Landau-Ginzburg theory and yields a maximal subgroup of the visible spacetime gauge group, which is $E_6$ for $N=10$, $SO(10)$ for $N=8$, and $SU(5)$ for $N=6$. There are also 16 left-moving free fermions which generate the hidden $E_8$.

If our Landau-Ginzburg theory is defined on a $\mathbb{Z}_m$-orbifold, then our supperpotential has the following property:
$$\mathcal{W}(\epsilon^{\omega_i}\Phi_i,\epsilon^{n_a}\Lambda^a)=\epsilon^m \mathcal{W}. $$ 
More explicitly, the $F_a(\Phi_i)$ are quasi-homogeneous polynomials of degree $d_a=m-n_a$.\\ In this case, our model has left and right $U(1)$ symmetry. First let us define:
$$q_i=\frac{\omega_i}{m}\quad, \quad q_a=\frac{n_a}{m}.$$
Then, the left and right $U(1)$ charges of the fields are stated in Table 1.   
\def\tablerule{\omit&\multispan{7}{\tabskip=0pt\hrulefill}&\cr}
\def\tablepad{\omit&height3pt&&&&&&&\cr}
$$\vbox{\offinterlineskip\tabskip=0pt\halign{
		\hskip.5in\strut$#$\quad&\vrule#&\quad\hfil $#$ \hfil\quad &\vrule #&\quad
		\hfil $#$
		\hfil \quad&\vrule #& \quad\hfil $#$ \hfil \quad&\vrule#&\quad $#$\cr
		&\omit&\hbox{Field}&\omit&\hbox{Left $U(1)$ charge $q$
		}&\omit&\hbox{Right $U(1)$ charge $\bar q$}&\omit&\cr
		\tablerule\tablepad
		&& \phi_{i} &&q_{i}&&q_{i}&&\cr
		\tablerule\tablepad
		&&\psi^{i}&&q_{i}&&q_{i}-1&&\cr
		\tablerule\tablepad
		&&\lambda^{a}&&q_{a}-1&&q_{a}&&\cr
		\tablerule
		\noalign{\bigskip}
		\noalign{\narrower\noindent{\bf Table 1:}
			Left and right $U(1)$ charges of fields in Landau-Ginzburg theory.}
}}$$
Combining the $\mathbb{Z}_m$ orbifold structure with $\mathbb{Z}_2$ orbifold structure, coming from Ramond (R) boundary condition or Neveu–Schwarz (NS) boundary condition on the fermions, we get $\mathbb{Z}_{2m}$ orbifold. So, we get $k=0,\cdots 2m-1$ sectors, where $k$ odd sectors belong to NS sector and $k$ even sectors belong to R sector. We have the following boundary conditions on the fields in $k$-th sector.
$$f(\tau,\sigma+2\pi)=e^{-iqk}f(\tau,\sigma),$$ 
where $q$ is the left charge of the fermionic field. Using these boundary conditions, we can expand our fields using Fourier series:
$$f(\tau,\sigma)=\sum_{r\in \mathbb{Z}-\frac{qk}{2\pi}}f_re^{ir\sigma},  $$       
where the $f_r$ are called the modes of the fields and after quantization $f_r$ will be operators. Using Born-Oppenheimer approximation for computing massless spectrum, we truncate the quantum theory the lowest modes. We demand the following commutation and anti-commutation relations on the lowest modes of the fields ($r>0$):
\begin{align*}
	&[\phi_r,\bar{\phi}_{r^\prime}]=\delta_{0,r+r^\prime}, \qquad[\bar{\phi}_r,\phi_{r^\prime}]=\delta_{0,r+r^\prime},\qquad [\bar{\phi}_0,\phi_0]=1, \\ 
	&\{\psi_r,\bar{\psi}_{r^\prime} \}=\delta_{0,r+r^\prime},\qquad \{\bar{\psi}_r,\psi_{r^\prime} \}=\delta_{0,r+r^\prime}, \qquad \{\psi_0,\bar{\psi}_0 \}=1. 	
	 \end{align*}   
	 In this case vacuum of the theory in each sector is annihilated by the following operators:
	 \begin{align*}
	 	\phi_r\Ket{0}=\bar{\phi}_r\Ket{0}=\bar{\phi}_0\Ket{0}=\psi_r\Ket{0}=\bar{\psi}_r\Ket{0}=\psi_0\Ket{0}=0. 
	 \end{align*}
	 To calculate massless spectrum, first we need to know what are the left and right charges and energy of the vacuum, $\Ket{0}_k$, in each sector. The following formulas give these information in the $k$-th sector.
	 \begin{align*}
	 	&q_k=-\sum_i q_i\Big(\frac{q_ik}{2}-[\frac{q_ik}{2}]-\frac{1}{2}\Big)-\sum_a (1-q_a)\text{Frac}\Big[\frac{1}{2}(1-q_a)k \Big], \\
	 	&\bar{q}_k=\sum_i (1-q_i)\Big(\frac{q_ik}{2}-[\frac{q_ik}{2}]-\frac{1}{2} \Big)+\sum_a q_a\text{Frac}\Big[\frac{1}{2}(1-q_a)k \Big], \\
	 	&E_k=-\frac{1}{2}\sum_i\Big(\frac{q_ik}{2}-[\frac{q_ik}{2}]-\frac{1}{2} \Big)^2+\frac{1}{2}\sum_a q_a\text{Frac}\Big[\frac{1}{2}(1-q_a)k \Big]^2+E_k^{(0)}
	 \end{align*}
	  where 
	  \begin{align*}
	  	E_k^{(0)}=\begin{cases}
	  	(3-\tilde{r})/8 \quad\quad\hspace*{.1 cm} \text{$k$ even} \\
	  	-5/8 \quad\quad\quad\quad \text{$k$ odd}
	  	\end{cases}
	  \end{align*}
	  and 
	 \begin{align*}
	 	\text{Frac}[x]=\begin{cases}
	 	1/2  \hspace*{2.5 cm} \text{integer}\hspace*{.15 cm} x \\
	 	x-[x]-1/2    \hspace*{.8 cm} \text{non-integer}\hspace*{.15 cm} x.
	 	\end{cases}
	 \end{align*}
We actually need this function to include all the cases in a closed form.

In finding massless sates, if energy level of the vacuum is negative, we can increase it, by acting negative modes on it. More explicitly, 
\begin{align*}
	L_0\Ket{0}=E_k\Ket{0}
\end{align*} 
then
\begin{align*}
L_0f_{-r}\Ket{0}=(E_k+r)\Ket{0}.
\end{align*}	  
In this way, we can construct states which are massless with respect to $L_0$. Also, to make sure that our states satisfy orbifold structure, we truncate these massless states to the states with half-integer left and right $U(1)$ charges. 

There is one more refinement in for theses states. These states should be also massless under $\bar{L}_0$ operator. For a $(0,2)$ supersymmetric theory, $\bar{L}_0$ can be written in the following form 
\begin{align*}
	\bar{L}_0=\{\bar{Q}_-,\bar{Q}_+\}, \qquad \bar{Q}_-^2=\bar{Q}_+^2=0.
\end{align*}
So, finding such states is equivalent to calculating $\bar{Q}_+$ cohomology. The explicit form of $\bar{Q}_+$ has been provided in \cite{Witten:1993jg,Fre:1992hp},
\begin{align*}
\bar{Q}_+=i\oint (i\bar{\psi}^i\bar{\partial}\phi_i+\mathcal{W}|_{\theta}).
\end{align*}
It has been in \cite{Kachru:1993pg} that, calculating above cohomology, is equivalent to calculating $\bar{Q}_{+,R}$ cohomology,
\begin{align*}
 \bar{Q}_{+,R}=i\oint i\bar{\psi}^i\bar{\partial}\phi_i,
\end{align*}
which simply removes all the states including right moving fermions, and then calculate $\bar{Q}_{+,L}$ cohomology,
\begin{align*}
	\bar{Q}_{+,L}=i \oint \mathcal{W}|_{\theta}=i\oint d\sigma \sum_a\lambda^aF_a(\phi_{i}).
\end{align*}  
The next step is to replace the fields by their lowest modes, which means
\begin{align*}
	f \rightarrow f_{-r}e^{-ir\sigma}+f_{-r+1}e^{i(-r+1)\sigma}
\end{align*}
Then in the above periodic integration over $\sigma$, only the terms survive that they have the sum of the indices of the fields are zero, because:
\begin{align*}
	\oint d\sigma e^{in\sigma}=\begin{cases}
	0     \qquad \hspace*{.2 cm} n\ne 0 \\
	2\pi   \qquad n=0.
	\end{cases}
\end{align*}
Having  $\bar{Q}_{+,L}$ explicitly in hand, we can calculate its cohomology, to do the last refinement on the states we have already found. To do so, we categorize the states by their charges. Let us call $V_{(q,\bar{q})}$ to be the vector space of the states with left and right $U(1)$ charges $(q,\bar{q})$. To see how many of these states survives after $\bar{Q}_{+,L}$ cohomology, consider the following sequence:
\begin{align*}
	\cdots \xrightarrow{} V_{(q,\bar{q}-1)} \xrightarrow{\bar{Q}_{+,L}^{(\star)}}  V_{(q,\bar{q})} \xrightarrow{\bar{Q}_{+,L}^{(\star\star)}}  V_{(q,\bar{q}+1)} \xrightarrow{}  \cdots
\end{align*}
Then the number of states surviving in  $V_{(q,\bar{q})}$, is equal to the dimension of $H_{(q,\bar{q})}$, 
\begin{align*}
	H_{(q,\bar{q})}=\text{Ker}\bar{Q}_{+,L}^{(\star\star)}\Big/\text{Im}\bar{Q}_{+,L}^{(\star)}
\end{align*}

\section{The program}
Our program is attached to the paper. To use our program, you need to have Mathematica installed in your computer. Then you need to install (or Get[$\cdots$/grassmann.m]) grassmann package in your Mathematica using the following link:

\url{http://people.brandeis.edu/~headrick/Mathematica/grassmann.m}

Then you need to specify your model by giving the following information to the program:
\begin{itemize}
	\item  \textbf{Nsector}: Total number of sectors which your orbifold contains, $2m$,
	\item  \textbf{k}: The sector that you are looking for massless states,
	\item  \textbf{NFreeFermions}: number of free left-moving fermions the model contains, N,
	\item  \textbf{ChargeOfFields}: Charge of chiral superfields $\omega_i$,
	\item $\textbf{F}_a$: The quasi-homogeneous polynomials appearing in the superpotential,
	\item  \textbf{DegreeOfPolynomials}: Degree of the quasi-homogeneous polynomials, $d_a$. 
\end{itemize} 

As an example, for the Quintic in \cite{Kachru:1993pg}, we have $10$ sectors with 
\begin{align*}
	W=\frac{1}{5}\Big(x_1^5+x_2^5+x_3^5+x_4^5+x_5^5\Big),
\end{align*} 
which the quasi-homogeneous polynomials are $F_a=\partial W/\partial x_a$ and the theory contains $10$ free fermions. If you are interested in finding massless spectrum for $k=1$ sector, then you need to insert the following to the input:
\begin{align*}
	&\text{NSector}=10;\\ 
	&\text{k}=1;\\
	&\text{NFreeFermions}=10;\\
	&\text{ChargeofFields}=\{1,1,1,1,1\};\\
	&\text{DegreeofPolynomials}=\{4,4,4,4,4\};\\
	&\text{F}_1=\text{x}_1^4;\\
	&\text{F}_2=\text{x}_2^4;\\
	&\text{F}_3=\text{x}_3^4;\\
	&\text{F}_4=\text{x}_4^4;\\
	&\text{F}_5=\text{x}_5^4;\\
\end{align*}
After evaluating the Mathematica notebook, the output will be in the following form:
\begin{align*}
	&\text{For Singlets:}\\
	&\text{We have 5 states with charges $(\hspace*{.1 cm}0\hspace*{.1 cm} ,\hspace*{.1 cm}-\frac{3}{2}\hspace*{.1 cm})$}\\
	&\text{We have 305 states with charges $(\hspace*{.1 cm}0\hspace*{.1 cm} ,\hspace*{.1 cm}-\frac{1}{2}\hspace*{.1 cm} )$}\\
	&\text{We have 101 states with charges $(\hspace*{.1 cm}2\hspace*{.1 cm} ,\hspace*{.1 cm}\frac{1}{2}\hspace*{.1 cm} )$}\\
	&\text{For 10's of SO(10): }\\
	&\text{We have 101 states with charges $(\hspace*{.1 cm}1 \hspace*{.1 cm},\hspace*{.1 cm}-\frac{1}{2}\hspace*{.1 cm})$}\\
	&\text{For Adjoint representation of SO(10): }\\
	&\text{We have 1 states with charges $(\hspace*{.1 cm}0\hspace*{.1 cm} ,\hspace*{.1 cm}-\frac{3}{2}\hspace*{.1 cm})$}\\
\end{align*} 
which agrees with the computations in \cite{Kachru:1993pg}. 

At the end, we need to emphasize that calculations in $k=0$ sector are time consuming and you might need to run it on a professional computer, depending on the complexity of your model.

\section{Acknowledgements}
We appreciate E.~Sharpe and L.~Anderson for providing the opportunity to do this project. We also would like to thank J.~Gray, I.~Melnikov, and A. Sarshar for useful discussions.

\end{document}